\documentclass[published]{JHEP}
\JHEP{07(2001)015}

\usepackage{amssymb,bbm}

\newcommand{\UU}{\mathop{\rm U}}
\newcommand {\n}{\nonumber \\}
\newcommand {\Tr}{\mbox{Tr\,}}
\newcommand {\ee}{\mbox{e}}
\newcommand {\dd}{\mbox{d}}
\newcommand {\del}{\partial}
\newcommand {\defeq}{\stackrel{\rm def}{=}}
\newcommand{\id}{\mathbbm{1}}

\title{Ginsparg-Wilson fermions in odd dimensions}

\author{Wolfgang Bietenholz\\
        Humboldt Universit\"{a}t, Institut f\"{u}r Physik\\
        Invalidenstr. 110, D-10115 Berlin, Germany\\
        E-mail: \email{bietenho@physik.hu-berlin.de}}

\author{Jun Nishimura\thanks{On leave of absence from Department
of Physics, Nagoya University, Nagoya 464-8602, Japan.}\\
        The Niels Bohr Institute\\ 
	Blegdamsvej 17, DK-2100 Copenhagen \O, Denmark\\
	E-mail: \email{nisimura@nbi.dk}}

\abstract{The Ginsparg-Wilson relation, if written in a suitable form,
can be used as a condition for lattice Dirac operators of massless
fermions also in odd dimensions.  The fermion action with such a Dirac
operator is invariant under a generalized parity transformation, which
reduces to the ordinary parity transformation in the (naive) continuum
limit.  The fermion measure, however, transforms non-trivially under
the generalized parity transformation, and hence the parity anomaly
arises solely from the fermion measure.  The analogy to the lattice
construction of chiral gauge theories in even dimensions is clarified
by considering a dimensional reduction.  We also propose a natural
definition of a lattice Chern-Simons term, which is consistent with
odd dimensional Ginsparg-Wilson fermions.}

\received{June 13, 2001}
\accepted{July 15, 2001\smash{\llap{\raise-1\baselineskip
		   \hbox{{\sc Revised: } July 11, 2001}}}}

\keywords{Field Theories in Lower Dimensions, Lattice Gauge Field Theories, Space-Time Symmetries, Anomalies in Field and String Theories}

\begin{document}

\section{Introduction}

Recently our understanding of chiral symmetry and gauge-field topology
on the lattice has developed considerably.  As an important
development, the so-called overlap formalism was
elaborated~\cite{NN}. Moreover, the Ginsparg-Wilson relation~\cite{GW}
was re-discovered~\cite{Has}.  This led in particular to a
gauge-invariant construction of abelian lattice chiral gauge
theories~\cite{Luescher_abelian} (see ref.~\cite{Golterman} for a
review on this subject).  In this paper, we show that the
Ginsparg-Wilson relation plays an important role also in odd
dimensions.

A single massless Dirac fermion in odd dimensions is parity invariant
on the classical level.  However, in order to regularize it in a gauge
invariant way, one has to break the parity invariance.  For example,
in the case of a Wilson fermion, the Wilson term breaks parity.  One
can obtain a massless fermion in the continuum limit by fine-tuning
the hopping parameter, but the breaking of parity remains in most
cases, which gives rise to the so-called parity
anomaly~\cite{parityanomaly}.

The Ginsparg-Wilson relation, if formulated in a suitable form, can be
used as a condition for lattice Dirac operators of massless fermions
in odd dimensions just as in even dimensions.  The fermion action with
such a Dirac operator is manifestly invariant under a generalized
parity transformation, which reduces to the ordinary parity
transformation in the naive continuum limit.  The fermion measure,
however, transforms non-trivially under the generalized parity
transformation.  Thus the parity anomaly arises solely from the
fermion measure in this formulation.

The connection to the even dimensional case can be clarified by first
considering a Ginsparg-Wilson fermion in $2n$ dimensions and then
performing a dimensional reduction down to $2n-1$ dimensions.  The
resulting system describes a \emph{doublet} of massless Dirac
fermions.  By projecting out one of these fermions, we obtain a single
odd-dimensional Ginsparg-Wilson fermion.  The projection can be done
in two different ways, leading to equivalent systems.  The first type
of projection can also be used for the Wilson fermion, but it has no
counterpart in even dimensions.  The second type of projection is
possible only for the Ginsparg-Wilson fermion and it is analogous to
the one used in constructing lattice chiral gauge theories in even
dimensions.  We study this analogy in detail and discuss in particular
the relation to the overlap formalism.  We also propose a natural
definition of a lattice Chern-Simons term, which is consistent with
odd dimensional Ginsparg-Wilson fermions.

For concreteness, we deal with three dimensions, which is relevant for
most applications, in particular for high temperature
superconductivity. However, the generalization to arbitrary odd
dimensions is straightforward.

The organization of this paper is as follows.  In section~\ref{GWFd3}
we introduce the Ginsparg-Wilson relation in odd dimensions.  We show
that any Dirac operator satisfying the Ginsparg-Wilson relation is
invariant under the generalized parity transformation.  In
section~\ref{dimred} we describe the dimensional reduction from ${\rm
D}=4$ to ${\rm D}=3$ and the standard projection to a single Dirac
fermion in the case of Wilson fermions.  In section~\ref{3dGW} we
apply the dimensional reduction to Ginsparg-Wilson fermions. In
section~\ref{g-p-trafo} we discuss the generalized parity invariance
in the context of dimensional reduction and clarify its relation to
the generalized chiral symmetry in even dimensions~\cite{ML}.  In
section~\ref{analog4D} we consider the analogy to the lattice
construction of chiral gauge theories in even dimensions.  In
section~\ref{overlap} we comment on the relation to some previous
results obtained from the overlap formalism.  In section~\ref{CS} we
propose a lattice Chern-Simons term, which follows naturally from the
previous considerations, and section~\ref{summary} is devoted to our
conclusions.

\section{The Ginsparg-Wilson fermion in odd dimensions}
\label{GWFd3}

The Ginsparg-Wilson relation in four dimensions
is commonly written as\footnote{The right-hand sides of eqs.~(\ref{GWR5}), (\ref{GWR4}),
(\ref{GWR3}) are understood as convolutions in c-space.}
\begin{equation} 
\label{GWR5}
D^{(4)} \gamma_{5} + \gamma_{5} D^{(4)} = a \, D^{(4)} \gamma_{5}
D^{(4)} \,,
\end{equation}
where $a$ is the lattice spacing.  It turned out that this condition
for the 4D lattice Dirac operator $D^{(4)}$ enables the physical
properties related to chirality to be represented correctly on the
lattice.  It is crucial for $D^{(4)}$ to be local (with exponentially
decaying correlations) and free of massless species doublers.
Assuming as usual ``$\gamma_{5}$-Hermiticity'' for $D^{(4)}$,
\begin{equation}  \label{gamma5H}
D^{(4) \dag}= \gamma_{5} D^{(4)} \gamma_{5} \,,
\end{equation}
we can re-write the Ginsparg-Wilson relation in a form, which does not
involve $\gamma_{5}$, and which is therefore compatible with odd
dimensions,
\begin{equation} 
\label{GWR4}
D^{(4)} + D^{(4) \dag} = a \, D^{(4) \dag} D^{(4)} \,.
\end{equation}
Hence we denote a 3D lattice Dirac operator $D^{(3)}$ as a
Ginsparg-Wilson operator if it obeys
\begin{equation} 
\label{GWR3}
D^{(3)} + D^{(3) \dag} = a \, D^{(3) \dag} D^{(3)} \,,
\end{equation}
and we expect it to describe a massless fermion in ${\rm D}=3$.  As in
even dimensions, a solution of this relation has to take the form
\begin{equation}
D^{(3)} = \frac{1}{a} \, (1 - V) \,,
\label{3dGWform}
\end{equation}
where the operator $V$ is \emph{unitary}.  In the naive continuum
limit, $V$ should turn into the identity operator.  Explicit solutions
for the 3D Ginsparg-Wilson operator are discussed in
section~\ref{overlap}.

The corresponding action
\begin{equation}
S = a^3 \sum _{x} \bar{\psi} (x) D^{(3)}(U) \psi (x) \,,
\label{action_projected}
\end{equation}
is invariant under
\begin{eqnarray}
&& \psi (x) \mapsto i \, {\cal R} \, V \, \psi (x) \,; \qquad
\bar{\psi} (x) \mapsto i \, \bar{\psi} (x) \, {\cal R} \,; \qquad
U_{\mu}(x)  \mapsto  U_{\mu}^{P}(x) \,,
\label{projected_parity} \\
{\rm or} && \psi (x) \mapsto i \, {\cal R} \, \psi (x) \,; \qquad
\bar{\psi} (x) \mapsto i \, \bar{\psi} (x) \, V \, {\cal R}\,; 
\qquad
U_{\mu}(x)  \mapsto  U_{\mu}^{P}(x) \,,
\label{projected_parityB}
\end{eqnarray}
where ${\cal R}$ is the space-time reflection operator, ${\cal R}: \ x
\rightarrow -x$.  $U^{P}$ is the parity transformed gauge
configuration,
\begin{equation}
U_{\mu}^{P}(x) = U_{\mu}(-x -a\hat \mu )^{\dag} \,,
\end{equation}
$\hat \mu$ being a unit vector in $\mu$ direction.  We assume that the
3D Ginsparg-Wilson operator $D^{(3)}$ has the property
\begin{equation}
D ^{(3)} (U) ^{\dag} = {\cal R } D ^{(3)}(U^P) {\cal R } \,,
\label{reflect}
\end{equation}
which holds also for the 3D Wilson operator.  If the right-hand side
of the relation~(\ref{GWR3}) would vanish --- as is the case with the
continuum massless Dirac operator $\sigma _\mu (\del _\mu + i A_\mu
(x) )$ --- then the action would be invariant under the ordinary
parity transformation, which is obtained from
eq.~(\ref{projected_parity}) or~(\ref{projected_parityB}) by replacing
$V$ by the identity operator.  The right-hand side of~(\ref{GWR3})
being non-zero, the action~(\ref{action_projected}) is not invariant
under the ordinary parity transformation, but it still has exact
invariance on the lattice under the generalized parity transformation
(\ref{projected_parity}) or (\ref{projected_parityB}), which reduces
to the ordinary one in the naive continuum limit.

On the other hand, the fermion measure does change non-trivially under
transformation~(\ref{projected_parity}) or~(\ref{projected_parityB}),
\begin{equation}  
\label{meastransf}
\dd \psi \, \dd \bar \psi \mapsto (\det V)^{-1} \, \dd \psi \, \dd
\bar \psi \,,
\end{equation}
which gives rise to the parity anomaly in this formulation.  This is
in contrast to the situation with the Wilson fermion, where the parity
breaking appears explicitly in the action through the Wilson term.  A
massless Wilson fermion is achieved only by fine-tuning the hopping
parameter.  On the other hand, for Ginsparg-Wilson fermions the pole
of the propagator at zero momentum is preserved to all orders in
perturbation theory, hence one obtains a single massless Dirac fermion
without fine-tuning.

The situation is analogous to the generalized chiral symmetry in even
dimensions~\cite{ML}.  Note that similar to the Nielsen-Ninomiya
theorem, a local, undoubled lattice fermion in odd dimensions cannot
obey the standard parity symmetry. This rule is circumvented by the
lattice modified parity invariance~(\ref{projected_parity})
resp.~(\ref{projected_parityB}).  In order to clarify the relation to
the even dimensional Ginsparg-Wilson fermions, we formally construct
$D^{(3)}$ from $D^{(4)}$ by means of dimensional reduction and a
subsequent projection.  In the next section, we consider these
operations in the case of Wilson fermions to establish the conceptual
basis.

In passing, we comment that if $D^{(3)}$ satisfies the Ginsparg-Wilson
relation, so does $D^{(3)\dag}$.  Therefore the action
\begin{equation}
S ' = -a^3 \sum _{x} \bar{\psi} ' (x) D^{(3)} (U)^{\dag} \psi ' (x)
\label{Sprime}
\end{equation}
is also expected to describe a massless Dirac fermion.  However, the
parity anomaly is represented by  $\det( V^\dag )$  instead of 
$\det V$.  This manifests the well-known regularization dependence of
parity anomaly.  We will discuss this issue in more detail in
section~\ref{overlap}.  We also note that the action~(\ref{Sprime})
can be transformed to the form~(\ref{action_projected}) by the field
re-definitions $\psi ' (x) = V \psi (x)$ and $\bar{\psi} ' (x) =
\bar{\psi} (x)$.  However, the measure is different, $ \dd \psi ' \,
\dd \bar {\psi} ' = (\det V)^{-1} \, \dd \psi \, \dd \bar \psi $.
Therefore, the two systems~(\ref{action_projected}) and~(\ref{Sprime})
with the standard measure are \emph{not} equivalent.\footnote{Note
that in ${\rm D}=4$ the corresponding two systems are equivalent due
to the $\gamma_{5}$-hermiticity~(\ref{gamma5H}). $D^{(3)}$, however,
has the property~(\ref{reflect}) instead, which makes a crucial
difference.}  In fact the two Dirac fermions form a parity doublet, as
we will see in section~\ref{3dGW}.

\section{Dimensional reduction of Wilson fermions}
\label{dimred}

We start from a 4D Wilson fermion with a lattice gauge field $U_\mu
(x)$ on a 4D periodic lattice $L_1 \times \cdots \times L_4$.  Then we
perform a ``dimensional reduction'' in the framework of the lattice
formulation to obtain a 3D system.  By ``dimensional reduction'', we
mean that we let $L_4 =1$ and take the gauge field to be $U_4 (x) = 1$
for any lattice site $x$. $L_{1} \dots L_{3}$ may be finite or
infinite.  The system we obtain is described by the action
\begin{equation}
S = a^3 \sum _{x} \bar \Psi (x) [ D_{\rm w}(U) + m ] \Psi (x) \,,
\label{actionWilson}
\end{equation}
where $D_{\rm w}$ is given by
\begin{equation} 
\label{3DWilson}
D_{\rm w} = \frac{1}{2} \sum _{\mu = 1} ^3 \{ \gamma _\mu (\nabla _\mu
^ * + \nabla _\mu ) + a \nabla _\mu ^ * \nabla _\mu \} \,,
\end{equation}
and the forward resp.\ backward lattice derivatives act as
\begin{eqnarray}
\nabla_{\mu} \Psi (x) &=& \frac{1}{a} \Big[ U(x,\mu ) \Psi (x+a\mu )
- \Psi (x) \Big] 
\nonumber\\
\nabla_{\mu}^{*} \Psi (x) &=& \frac{1}{a} \Big[ \Psi (x) -
U(x-a\hat \mu )^{-1} \Psi (x-a \hat \mu ) \Big] \,.
\nonumber
\end{eqnarray}
For the 4D gamma matrices we use the representation
\begin{equation}
\gamma _\mu =\pmatrix{0  & \sigma _\mu  \cr
\sigma _\mu  &  0};\qquad 
\gamma _4 =\pmatrix{0  &  i \id  \cr
- i \id  &  0};\qquad 
\gamma _5 =\pmatrix{\id  & 0  \cr
0   &  - \id}.
\end{equation}
Note that $D_{\rm w}$ inherits $\gamma_{5}$-Hermiticity from the 4D
Wilson operator,
\begin{equation} 
\label{g5H4d}
D_{\rm w}   (U)^\dag
=    \gamma _5 D_{\rm w} (U) \gamma _5  \,.
\end{equation}

In addition, $D_{\rm w}$ has some properties which the 4D Wilson
operator does not have. Firstly,
\begin{equation}
D_{\rm w}(U) \, \Gamma = \Gamma \,  D_{\rm w}(U) \,,
\label{ordinary_decoupling}
\end{equation}
\cite{NPRL} where $\Gamma$ is a unitary and hermitean matrix defined
as
\begin{equation}
\Gamma = i \gamma _4 \gamma_5 = \pmatrix{0  &  \id  \cr
\id  &  0} .
\end{equation}
Due to commutation relation~(\ref{ordinary_decoupling}), the
action~(\ref{actionWilson}) is invariant under
\begin{equation}
\Psi (x) \mapsto  \ee ^{i\alpha \Gamma } \Psi (x)\,;\qquad
\bar{\Psi} (x) \mapsto  \bar{\Psi} (x) \ee ^{- i\alpha \Gamma } \,;\qquad
U_{\mu}(x)  \mapsto  U_{\mu}(x) \,,
\label{Gamma_inv}
\end{equation}
where $\alpha$ is a real parameter.  Under space-time reflection,
$D_{\rm w}$ behaves as eq.~(\ref{reflect}),
\begin{equation}
D_{\rm w} (U) ^{\dag} = {\cal R } D_{\rm w} (U^P) {\cal R } \,.
\label{parity_rel}
\end{equation}

Using the relations~(\ref{g5H4d}) and~(\ref{parity_rel}), we see that
the action~(\ref{actionWilson}) is invariant under the (standard)
parity transformation\footnote{In
transformation~(\ref{ordinary_parity}), $\gamma_5$ could be replaced
by $\gamma _4$.  More generally, one can replace $\gamma _5$ by
$\gamma _5 \ee ^{i \alpha \Gamma}$, as we see if we combine the
transformations~(\ref{Gamma_inv}) and~(\ref{ordinary_parity}).
However, we stay with~(\ref{ordinary_parity}), because the further
generalizations do not provide additional insight.}
\begin{eqnarray}
\Psi (x)& \mapsto & \gamma _5 \Psi (-x)  = \gamma_5 {\cal R} \Psi (x) \n
\bar{\Psi} (x)& \mapsto & \bar{\Psi} (-x) \gamma _5 
= \bar{\Psi} (x) {\cal R} \gamma_5 \n
U_{\mu}(x) & \mapsto & U_{\mu}^{P}(x) \,.
\label{ordinary_parity}
\end{eqnarray}

Due to property~(\ref{ordinary_decoupling}), the
action~(\ref{actionWilson}) naturally decouples into two Dirac
fer\-mions, where ${1}/{2}(1 \pm \Gamma)$ acts as a projection
operator.  Let us introduce the unitary matrix $\bar U$ that
diagonalizes $\Gamma $,
\begin{equation}
\bar U ^\dag \, \Gamma \, \bar U = \pmatrix{\id  &  0 \cr
0  &  -\id}= \gamma_5\,;\qquad
\bar U = \frac{1}{\sqrt{2}}
\pmatrix{\id  &  \id  \cr
\id   &  - \id}. 
\label{Udef}
\end{equation}
Using this unitary matrix $\bar U$, also the Wilson operator can be
brought into a block-diagonal form as
\begin{equation}
\bar U ^\dag  D_{\rm w} \bar U = \pmatrix{D_{\rm w} ^{(3)}  &  0 \cr
 0   &  D_{\rm w} ^{(3)\dag}},
\label{blockDw}
\end{equation}
which follows only from the properties~(\ref{g5H4d})
and~(\ref{ordinary_decoupling}).  From the explicit
form~(\ref{3DWilson}), we see that $D_{\rm w} ^{(3)}$ in
eq.~(\ref{blockDw}) is given by the 3D Wilson operator
\begin{equation}
D_{\rm w} ^{(3)} = \frac{1}{2} \sum _{\mu = 1} ^3 \{ \sigma _\mu
(\nabla _\mu ^ * + \nabla _\mu ) + a \nabla _\mu ^ * \nabla _\mu \} \,.
\label{3DWDop}
\end{equation}
By decomposing the fermion fields as
\begin{equation}
\Psi(x) = \bar{U}\pmatrix{\psi(x)  \cr
\psi ' (x)},\qquad
\bar{\Psi}(x) =
(\bar{\psi} (x) , \bar{\psi} ' (x) ) \, \bar{U} ^\dag \,,
\label{decomposedfermion}
\end{equation}
we find that the action~(\ref{actionWilson}) decouples into two 3D
Dirac fermions, with mass and Wilson parameter of opposite signs.
Note also that the integration measure decouples simply as
\begin{equation}
\dd \Psi \, \dd \bar{\Psi} = \dd \psi \, \dd \bar{\psi}
\cdot \dd \psi ' \, \dd \bar{\psi} ' \, ,
\label{Jacobian}
\end{equation}
without a nontrivial jacobian factor.  The unitary transformation
performed here corresponds to a change of the representations of the
4D $\gamma$ matrices.

One can extract one of the two 3D Dirac fermions by imposing the
constraint
\begin{equation}
\Gamma \Psi = \Psi\,;\qquad
\bar{\Psi} \Gamma  = \bar{\Psi} \,.
\label{ordinary_projection}
\end{equation}
However, this constraint is not compatible with the parity
transformation~(\ref{ordinary_parity}), since $\Gamma$ does not
commute with $\gamma_5$.  Therefore, parity
invariance~(\ref{ordinary_parity}) does \emph{not} survive the
projection~(\ref{ordinary_projection}).  This can be understood
intuitively since the parity transformation~(\ref{ordinary_parity})
exchanges the two 3D Dirac fermions, hence it cannot be described by
using only one of them.

We also remark that the existence of the (standard) parity
invariance~(\ref{ordinary_parity}) in the 3D \emph{doublet} system does
not imply the masslessness of the two Dirac fermions.  It only implies
that they have masses of opposite signs with the same magnitude.  It
is for a \emph{single} 3D Dirac fermion that the parity invariance on
the classical level implies masslessness.

\section{3D Ginsparg-Wilson fermions from dimensional reduction}
\label{3dGW}

Let us now consider a 4D Ginsparg-Wilson fermion and perform a
dimensional reduction to ${\rm D}=3$.  The system we obtain is
described by the action
\begin{equation}
S = a ^ 3 \sum _{x} \bar{\Psi} (x) D(U) \Psi (x) \,,
\label{actionGW}
\end{equation}
where $D$ satisfies the Ginsparg-Wilson relation
\begin{equation}
D + D^{\dagger} = a \, D^{\dagger} D \,.
\label{GWrelation}
\end{equation}
We also assume that it fulfills the
properties~(\ref{g5H4d}),~(\ref{ordinary_decoupling})
and~(\ref{parity_rel}).

Since $D$ has all the properties of $D_{\rm w}$ that we used in
section~\ref{dimred}, all the statements in the previous Section apply
to the present case.  Hence the system~(\ref{actionGW}) is invariant
under the (standard) parity transformation~(\ref{ordinary_parity}) and
it decouples into two Dirac fermions if we apply $\frac{1}{2}( 1 \pm
\Gamma)$ as a projector.  We can bring the operator $D$ into a
block-diagonal form as
\begin{equation}
\bar U ^\dag \, D \, \bar U = \pmatrix{D ^{(3)}  & 0  \cr
 0   &  D ^{(3)\dag}}.
\label{D3}
\end{equation}
Due   to   the   3D   Ginsparg-Wilson  relation   for   the   doublet,
eq.~(\ref{GWrelation}),  the  operator  $D ^{(3)}$  in  eq.~(\ref{D3})
obeys the 3D  singlet Ginsparg-Wilson relation~(\ref{GWR3}).  By using
the    decomposition~(\ref{decomposedfermion}),    we    obtain    the
systems~(\ref{action_projected}) and~(\ref{Sprime}) with the standard
integration measure.  Therefore, the system~(\ref{actionGW}) describes
a doublet  of 3D Ginsparg-Wilson  fermions, which are  exchanged under
the standard  parity transformation~(\ref{ordinary_parity}).   One can
extract one  of the  two 3D Ginsparg-Wilson  fermions by  imposing the
constraint~(\ref{ordinary_projection}).

In analogy to the even dimensional case~\cite{Luescher_abelian}, we
define the operator $\hat{\gamma}_5$,
\begin{equation}
\hat{\gamma}_5 = \gamma_5 (1 - a D) \,,
\label{gamhatdef}
\end{equation}
which is hermitean,
\begin{equation} \label{Hg5}
\hat{\gamma}_5 ^\dag = \hat{\gamma}_5 \,,
\end{equation}
due to the $\gamma_{5}$-hermiticity of $D$.  The Ginsparg-Wilson
relation~(\ref{GWrelation}) further implies
\begin{eqnarray}
\label{gamhatsquare}
(\hat{\gamma}_5)^2 &=& 1 \,, 
\nonumber\\
\gamma_5 D &=& - D \hat{\gamma}_5 \label{gamhatproj} \,.
\end{eqnarray}
Eqs.~(\ref{Hg5}) and~(\ref{gamhatsquare}) show that $\hat{\gamma}_5$
is also unitary.  In addition, eq.~(\ref{gamhatproj}) implies
\begin{equation}
\ee ^{i \alpha \gamma _5 } D \ee ^{i \alpha \hat{\gamma} _5 } = D \, ,
\qquad \ee ^{i \beta \hat{\gamma} _5 } D^\dag \ee ^{i \beta \gamma _5
} = D ^\dag \,,
\label{latticechiral}
\end{equation}
which is analogous to the generalized lattice chiral symmetry in even
dimensions~\cite{Luescher_abelian}.

In the present 3D case, the operator $\hat{\gamma}_5$ has the
additional property
\begin{equation}
\hat \gamma_{5} \, \Gamma = - \Gamma \, \hat \gamma_{5} \,,
\label{gamhatGam}
\end{equation}
which follows from relation~(\ref{ordinary_decoupling}).
Eqs.~(\ref{Hg5}),~(\ref{gamhatsquare}) and~(\ref{gamhatGam}) imply
that $\hat{\gamma}_5$ has the structure
\begin{equation}
\hat{\gamma}_5 
=\pmatrix{X  &&  i Y  \cr
- i Y  &&  -X} ,
\label{structure}
\end{equation}
where $X$, $Y$ are hermitean matrices satisfying  $X^2 + Y^2 = 1 $ 
and  $XY = YX$.  Solving~(\ref{gamhatdef}) for $D$, one obtains
\begin{equation}
D = \frac{1}{a} \pmatrix{1-X &   - i Y \cr
- i Y    &  1 - X}, 
\label{solveD}
\end{equation}
and eq.~(\ref{D3}) implies
\begin{equation}
D^{(3)} = \frac{1}{a} \, \{1 - (X + iY)\} \,.
\label{VDrel}
\end{equation}
Thus the 3D Ginsparg-Wilson operator $D^{(3)}$ takes the general
form~(\ref{3dGWform}) with the unitary operator $V$ given by $V = X +
iY$.

\section{Generalized parity transformations}
\label{g-p-trafo}

As in the case of the Wilson fermion, the
projection~(\ref{ordinary_projection}) is not compatible with the
parity transformation~(\ref{ordinary_parity}).  However, we will see
that the action~(\ref{actionGW}) is actually invariant under a more
general class of parity transformations due to the Ginsparg-Wilson
relation. Then we are going to show that there exist generalized
parity transformations, which \emph{are} compatible with the
projection~(\ref{ordinary_projection}).  In the next section, we will
see that again due to the Ginsparg-Wilson relation, there exists a
different type of projection, which is compatible with the standard
parity transformation~(\ref{ordinary_parity}).

Let us start from the ansatz
\begin{eqnarray}
\Psi (x) & \mapsto & {\cal R} A \Psi (x) \n
\bar{\Psi} (x) & \mapsto & \bar{\Psi} (x) B {\cal R}  \n
U_{\mu}(x) & \mapsto & U_{\mu}^{P}(x) \,,
\label{new_parity}
\end{eqnarray}
for a generalized parity transformation, where $A$ and $B$ are
operators acting on the fermion field.  Requiring the
action~(\ref{actionGW}) to be invariant under
transformation~(\ref{new_parity}), and using the
property~(\ref{parity_rel}), we obtain the condition
\begin{equation}
B D^\dag A = D \,.
\label{ABcond}
\end{equation}
The ordinary parity transformation~(\ref{ordinary_parity}) corresponds
to the particular case $A = B = \gamma_5$, which indeed satisfies the
condition~(\ref{ABcond}), cf.\ relation~(\ref{g5H4d}).

Exploiting the properties~(\ref{latticechiral}),
we obtain a more general solution to the condition~(\ref{ABcond}) 
as
\begin{equation}
A = \ee ^ {i\beta \gamma_{5}} \gamma_5 \ee ^{i\alpha \hat \gamma_{5}}\,,
\qquad B = \ee ^ {i\alpha \gamma_{5}} \gamma_5 \ee ^{ i\beta \hat
\gamma_{5}} \,,
\label{ABtrafo}
\end{equation}
with real parameters $\alpha$, $\beta$.  The invariance of the
projection~(\ref{ordinary_projection}) under the
transformation~(\ref{new_parity}) requires that the operators $A$ and
$B$ commute with $\Gamma$.  Such a set of operators $A$, $B$ can be
obtained by choosing  $\alpha = {\pi}/{2}$, $\beta = 0$  or 
$\alpha = 0$, $\beta = {\pi}/{2}$  in the generalized
solution~(\ref{ABtrafo}).  These two choices correspond to
\begin{eqnarray}
{\rm (i)}&\qquad& A = i \gamma_{5} \hat \gamma_{5} \,, \quad B = i \,, 
\label{ABsurvive1}\\
{\rm (ii)}&\qquad& A = i \,, \quad B =  i \gamma_{5} \hat \gamma_{5} \,, 
\label{ABsurvive2}
\end{eqnarray}
respectively.  Correspondingly, the projected fermion fields $\psi(x)$
and $\bar{\psi}(x)$ transform as~(\ref{projected_parity})
resp.~(\ref{projected_parityB}) under the generalized parity
transformation~(\ref{new_parity}).

Hence the existence of the generalized parity
transformation~(\ref{projected_parity}) or~(\ref{projected_parityB})
in the action~(\ref{action_projected}) can be understood in analogy to
the generalized chiral symmetry in even dimensions.

\section{Analogy to chiral gauge theories in ${\rm D}=4$}
\label{analog4D}

Due to property (\ref{gamhatproj}), the system (\ref{actionGW}) also
allows for another projection by imposing the constraint
\begin{equation}
\hat{\gamma}_5 \Psi = \Psi\,; \qquad \bar{\Psi} \gamma_5 = -
\bar{\Psi} \,,
\label{gammahat_projection}
\end{equation}
instead of the constraint~(\ref{ordinary_projection}).  This is in
fact the projection used to construct lattice chiral gauge theories in
the even dimensions~\cite{Luescher_abelian}.

The projection~(\ref{gammahat_projection}) is actually compatible with
the standard parity transformation~(\ref{ordinary_parity}), since
\begin{equation}
\hat{\gamma}_5 (U^P) \, ( {\cal R} \gamma _5 ) = ( {\cal R} \gamma _5
) \, \hat{\gamma}_5 (U) \,.
\end{equation}
The parity transformation written in terms of the projected fields
$\psi (x)$ and $\bar{\psi} (x)$ is given by eq.~(\ref{projected_parity}).

Next we note that eq.~(\ref{gamhatGam}) implies that $\hat{\gamma}_5$
has an equal number of eigenvalues $+1$ and $-1$.  Consequently, the
counterpart of the index in ${\rm D}=4$ vanishes identically,
\begin{equation}
\Tr \left( \frac{ 1 + \gamma_5 }{2}\right) - \Tr \left( \frac{ 1 -
\hat{\gamma}_5 }{2}\right) = \frac{1}{2} \Tr (\gamma_5 D) = 0 \,,
\end{equation}
where the trace $\Tr$ is taken over all the spinor, space-time and
gauge indices. Hence it is possible to diagonalize $\hat{\gamma}_5$,
for instance as
\begin{equation}
\tilde{U} ^\dag  \, \hat{\gamma}_5 \, \tilde{U}  
=\pmatrix{ \id  &  0 \cr 0  &  -\id}
= \gamma_5 \,,
\qquad \tilde{U} = 
\bar{U} \,
\pmatrix{1  &  0  \cr 0  &  V}
\, \bar{U} ^\dag \,
\pmatrix{1  &  0  \cr 0  &  - V^\dag}.
\label{Uhatdef}
\end{equation}
Using this matrix, we can bring the operator $D$ into a block-diagonal
form,
\begin{equation}
\Gamma  \, D \, \tilde{U} = 
\pmatrix{D ^{(3)}  & 0  \cr  0   &  D ^{(3)\dag}}.
\label{D3_2}
\end{equation}
By decomposing the fermion fields as
\begin{equation}
\Psi(x) = \tilde{U}\pmatrix{\psi(x)  \cr
\psi ' (x)},\qquad
\bar{\Psi}(x) =
(\bar{\psi} (x) , \bar{\psi} ' (x) ) \, \Gamma \,,
\label{decomposedfermion2}
\end{equation}
we see that the action~(\ref{actionGW}) decouples into two 3D
Ginsparg-Wilson fermions.  Note also that the integration measure
decouples as in eq.~(\ref{Jacobian}), without a nontrivial jacobian
factor, since  $\det \Gamma \det \tilde{U} = 1$.  Although the
decomposition~(\ref{decomposedfermion2}) is different
from~(\ref{decomposedfermion}), the resulting systems are
equivalent.

In ref.~\cite{Real}, the 4D Weyl fermion belonging to the real
representation of the gauge group has been studied using the
Ginsparg-Wilson fermion.  There, it was shown that the fermion measure
can be defined in a gauge invariant way, although the construction is
quite different from the one presented here.

\section{Relation to the overlap formalism}
\label{overlap}

So far, we have not specified a particular form of the Ginsparg-Wilson
operators.  In even dimensions, an explicit example of the
Ginsparg-Wilson operator $D^{(4)}$, which satisfies the
Ginsparg-Wilson relation (\ref{GWR5}) and other required properties,
is given by the overlap operator~\cite{overlapDirac}.  Up to moderate
coupling strength, the overlap operator is local and free of doubling
and mass renormalization\footnote{The collapse of the latter two
properties was observed in a strong coupling expansion~\cite{BS}.  For
instance, in quenched QCD$_{4}$ the applicability of the overlap
formula --- starting from $D_{\rm w}^{(4)}$ --- extends down to about
$\beta \approx 5.6$~\cite{China}.}.  The operator $D$, obtained by
dimensional reduction of an overlap operator $D^{(4)}$, is given by
\begin{equation}
D = \frac{1}{a} \left\{ 1 - {A_{\rm w} }/ 
{\sqrt{ A_{\rm w} ^ \dag A_{\rm w} }} \right\},\qquad
A_{\rm w} =  1 - a D_{\rm w}  \,,
\label{overlapDirac}
\end{equation}
where the Wilson operator $D_{\rm w}$ is defined in
eq.~(\ref{3DWilson}).  As in the even dimensional
case~\cite{Golterman,equivalence}, the construction presented in
section~\ref{analog4D} becomes mathematically isomorphic to the
overlap formalism~\cite{NN}.

The overlap formalism, which was originally proposed to construct a
lattice chiral gauge theory,\footnote{See, however,
ref.~\cite{IzubuchiNishimura}, where the problem of the so-called
Wigner-Brillouin phase choice proposed in ref.~\cite{NN} has been
clarified.}  has also been applied to odd
dimensions~\cite{NarayananNishimura,KikukawaNeuberger}.  In
ref.~\cite{NarayananNishimura} it was pointed out that the overlap
formalism provides a lattice regularization of a single 3D Dirac
fermion with manifest parity invariance except for the phase of the
fermion determinant.  This is due to the fact that the auxiliary
many-body hamiltonian, which plays the central role in the overlap
formalism, has a manifest parity invariance.  Explicitly, the
many-body hamiltonian is given by
\begin{equation}
{\cal H}(U) = \sum_{x} \, (\xi(x) ^\dag , \eta(x) ^\dag ) \, 
 H(U) \pmatrix{\xi(x)  \cr \eta(x)},
\label{manybody}
\end{equation}
with the hermitean operator
\begin{equation}
H = \gamma _5 \, (1 - a \, D_{\rm w} ) \, .
\end{equation}
The many-body hamiltonian (\ref{manybody}) 
has a manifest parity invariance
\begin{equation}
\xi (x) \mapsto  i \, \xi(-x)\,;\qquad
\eta (x) \mapsto  i \, \eta(-x)\,;\qquad
U_{\mu}(x)  \mapsto  U_{\mu}^{P}(x) \,.
\label{overlap_paritytr}
\end{equation}
In fact one can re-write the overlap formalism in a path integral form
following e.g.\ ref.~\cite{KikukawaYamada} to arrive at the formalism
in section~\ref{analog4D}.  Then one observes that the parity
transformation~(\ref{overlap_paritytr}) maps to the standard parity
transformation~(\ref{ordinary_parity}), which survives the
projection~(\ref{gammahat_projection}).

For the particular choice~(\ref{overlapDirac}), the corresponding 3D
Ginsparg-Wilson operator $D^{(3)}$ defined through~(\ref{D3})
or~(\ref{D3_2}) takes the form
\begin{eqnarray}
\label{overlapDirac3}
D ^{(3)} &=& \frac{1}{a} \left\{
1 - {A_{\rm w}^{(3)}}/{ \sqrt{A_{\rm w}^{(3)\dag} 
A_{\rm w}^{(3)}}} \right\} 
\\
A_{\rm w}^{(3)} &=& 1 - a D_{\rm w}^{(3)}  \,.
\label{overlapA3}
\end{eqnarray}
This coincides with the overlap Dirac operator first dervived by
Kikukawa and Neuberger~\cite{KikukawaNeuberger}.  The phase of the
effective action  $\det (D ^{(3)})$  is parity odd, and it is half
of the phase of  $\det (A_{\rm w}^{(3)})$.  The latter has been
calculated in the continuum limit~\cite{CosteLuescher}, and it turned
out that the correct parity anomaly --- described by the Chern-Simons
term --- is reproduced.

In general, infinitely heavy Dirac fermions in odd dimensions do not
decouple completely. They leave a remnant in the effective
action,\footnote{This is not too surprising, since the remnant
quantity appears in terms of a local operator.}  which can by
expressed again by a Chern-Simons term.  When one regularizes a single
massless Dirac fermion, one has to introduce a number of heavy
fermions with mass of the cutoff scale, such as Pauli-Villars
regulators or massive doublers for the Wilson fermion.  The parity
anomaly represented by the induced Chern-Simons term therefore depends
on the number of heavy fermions introduced in the regularized theory.
The coefficient of the Chern-Simons term has an ambiguity labelled by
an integer, which specifies a universality class.  In particular,
ref.~\cite{CosteLuescher} considers the effective gauge action after
integrating out the fermion fields, $\Gamma [A]$. In the limit of
infinite fermion mass it amounts to
\begin{equation}
^{\rm lim}_{m \to \infty} \, \Gamma [A] = 2 n \pi i e^{2}
I [A] \,,
\end{equation}
where $e$ is the charge, $I[A]$ is the Chern-Simons term, and $n \in Z
\!\!\! Z$. It is related to the massless limit as
\begin{equation}
^{\rm lim}_{m \to 0} \, {\rm Im} \, \Gamma [A] = (2n + 1) \pi e^{2} I
[A] + \pi h[A] \,,
\end{equation}
where $h[A]$ is an integer, which vanishes for smooth gauge
fields. The integer $n$ labels the universality classes.  For the
Wilson fermion, they were first studied in ref.~\cite{So}.  The above
choice~(\ref{overlapDirac3}) with~(\ref{overlapA3}) reproduces one of
these classes.

A natural question is whether we can also reproduce other classes by
using Ginsparg-Wilson fermions.  At this point, we remark that the
overlap solution to the Ginsparg-Wilson relation can be generalized in
two respects~\cite{EPJC}: first, the relation~(\ref{GWR3}) itself can
be generalized to
\begin{equation}
D^{(3)} + D^{(3)\dagger} = 2a D^{(3)\dagger} R D^{(3)} \,,
\end{equation}
where the kernel $R$ has to be local and it must not be
parity-odd.\footnote{As in even dimensions, also perfect~\cite{GW} and
classically perfect~\cite{Hasenfratz:1998ri} actions are solutions to
the Ginsparg-Wilson relation.  In their construction, the term
$R^{-1}$ appears in the renormalization group transformation term, and
the standard choice $R_{x,y}=\frac{1}{2}\delta_{x,y}$ is optimal for
locality of the free fermion~\cite{WBUJW}.}  This relation is solved by the generalized
overlap formula
\begin{equation}
D^{(3)} = \frac{1}{2a} \frac{1}{\sqrt{R}} \left( 1 - {A^{(3)}}/{ \sqrt{
A^{(3)\dagger} A^{(3)}}} \right) \frac{1}{\sqrt{R}}\,, \qquad A^{(3)}
= 1 - 2 a \sqrt{R} D_{{\rm w}}^{(3)} \sqrt{R} \,.
\end{equation}
In addition, we also have much freedom in the choice of $A^{(3)}$:
the Wilson operator $D_{{\rm w}}^{(3)}$ in its definition can be
replaced by many other lattice Dirac operators $D_{0}^{(3)}$.  If
$D^{(3)}_{0}$ is local and free of massless doublers, we arrive at a
sensible Ginsparg-Wilson operator $D^{(3)}$. (However, $D^{(3)}_{0}$
does not need to be parity invariant in any sense.)

If we now vary either $R$, or $D^{(3)}_{0}$ or both, we can change the
universality class. In fact, \emph{all} classes are accessible in this
way, so the set of 3D Ginsparg-Wilson fermions covers all of them. To
demonstrate this, we consider as an example the form
\begin{equation}
A^{(3)} = ( 1 - a D_{\rm w}^{(3)} ) \, (1 - 2 a \, D_{\rm
w}^{(3)})^{2n} \,, \qquad R_{x,y} = \frac{1}{8n+2} \, \delta_{x,y}
 \,,
\end{equation}
where $n = 0,1,2, \dots$ The phase of  $\det (A ^{(3)})$  for these
cases has been calculated in the continuum limit~\cite{CosteLuescher},
and the resulting parity anomaly is $(2n+1)$ times the one obtained
for the standard choice~(\ref{overlapA3}). Moreover, we can switch
from $D^{(3)}$ to $-D^{(3)\dag}$ (cf. Section 2), which is a
Ginsparg-Wilson operator for a kernel $R$ with opposite sign. This
changes also the sign of the parity anomaly.  Therefore, by exploiting
the freedom in choosing $R$ and $D_{0}$, we obtain Ginsparg-Wilson
fermions in all the universality classes --- labelled by an integer
ranging from $- \infty$ to $+ \infty $ --- which were identified
before for Wilson fermions~\cite{CosteLuescher}.

This confirms that we are not dealing with an anomaly in the usual
sense, where the amount is the same for all correct
regularizations. In the case of the parity anomaly, an infinite number
of universality classes coexist on the same level.

\section{A lattice formulation of the Chern-Simons term}
\label{CS}

Finally, the parity anomaly obtained from the 3D Ginsparg-Wilson
fermion suggests that
\begin{equation}
\ee ^{i S_{\rm CS}} \defeq 
\frac{\det (A_{\rm w}^{(3)})}{|\det (A_{\rm w}^{(3)})|} \, ,
\label{latticeCS}
\end{equation}
where we use $A_{\rm w}^{(3)}$ as defined in eq.~(\ref{overlapA3}),
can be used as a natural definition of the Chern-Simons term on the
lattice.  Of course, a more general operator $A^{(3)}$ can be used, as
long as we stay in the same universality class as the one obtained for
$A_{\rm w}^{(3)}$ in a given gauge configuration.

The normalization of $S_{\rm CS}$ is chosen in such a way that
it transforms as
\begin{equation}
S_{\rm CS} \mapsto S_{\rm CS} + 2 \pi \nu \,,
\end{equation}
under a gauge transformation, where $\nu$ is the winding number
characterizing the gauge transformation.  Consequently, the phase
factor $\ee ^{i S_{\rm CS}}$ should be gauge invariant, and indeed the
right-hand side of eq.~(\ref{latticeCS}) is manifestly gauge
invariant.  Since the index theorem holds in even dimensions on the
lattice for Ginsparg-Wilson fermions~\cite{Hasenfratz:1998ri}, one may
expect that the above definition captures the topological aspects of
gauge theories.\footnote{The term in eq.~(\ref{latticeCS}) is not
defined if the spectrum of $A^{(3)}$ includes a zero eigenvalue.  Note
that in this case, the Dirac operator $D^{(3)}$ (given in
eq.~(\ref{overlapDirac3})) is not defined either. However, the set of
configurations where this happens (the boundaries of topological
sectors) is statistically negligible~\cite{overlapDirac}.}  It is an
interesting question to study this term in the light of the No Go
theorem established in ref.~\cite{BDS}.  Finally, we note that a
related Chern-Simons lattice term has been suggested in
ref.~\cite{AK}.

\section{Discussion and outlook}
\label{summary}

In this paper, we have introduced the Ginsparg-Wilson relation in odd
dimensions and we revealed its importance.  We have shown that an
action based on any lattice Dirac operator satisfying the
Ginsparg-Wilson relation is invariant under a generalized parity
transformation.  The fermion measure, however, breaks parity
invariance, giving rise to the parity anomaly.  This is analogous to
the chiral symmetry in even dimensions: there the chiral anomaly
arises from the measure in the case of Ginsparg-Wilson
fermions~\cite{ML}.  Thus, we naturally realize Fujikawa's approach to
anomalies~\cite{Fujikawa} at a fully regularized level on the lattice.
This is in contrast to the situation with the Wilson fermion, where
the parity breaking appears explicitly in the action through the
Wilson term.

We note that the way the parity anomaly appears is similar to the
situation discussed in ref.~\cite{nonlocalparity}, where a single
Dirac fermion is considered in the continuum with Pauli-Villars
regularization.  After integrating out the Pauli-Villars regulator, a
(nonlocal) field re-definition of the fermion field is performed to
make the action parity invariant.  The parity anomaly arises from the
jacobian of the field re-definition.

Let us finally give some possible directions for future studies.  We
can impose a Majorana condition on the 3D Ginsparg-Wilson fermion to
obtain a massless 3D Majorana fermion, which can be used to construct
supersymmetric ${\cal N}=1$ Yang-Mills theory in three dimensions on
the lattice.  This provides a considerable simplification to the idea
presented in ref.~\cite{3dsusy} using the overlap formalism.  In
ref.~\cite{CPTanomaly}, it was suggested that parity anomaly in odd
dimensions may be related to a CPT anomaly in chiral gauge theories in
even dimensions.  Indeed, according to ref.~\cite{KlinkNishi} a CPT
anomaly exists in the exact solution~\cite{IzubuchiNishimura} of
two-dimensional $\UU(1)$ chiral gauge theories.  We hope that our
results may also shed light on that issue.  We have further proposed a
possible definition of the Chern-Simons term on the lattice.  It turns
into the correct form in the continuum limit.  As the index theorem
provides a sensible definition of the topological charge on the
lattice for Ginsparg-Wilson fermions, it is conceivable that the
proposed definition of the Chern-Simons term captures the topological
aspect of gauge theory.

To conclude, we note that the consistency of the Ginsparg-Wilson
fermion revealed in this paper provides further support for the new
approach to lattice chiral gauge theories
and extends the range of its application to odd dimensions.

\acknowledgments

J.N.\ would like to thank F.R.\ Klinkhamer for discussions on the CPT
anomaly, which motivated this work.  We further thank Y. Kikukawa and
P.~Sodano for useful comments.


\begin{thebibliography}{99}


\bibitem{NN}
R.~Narayanan and H.~Neuberger, \emph{A construction of lattice chiral
  gauge theories}, \npb{443}{1995}{305} [\hepth{9411108}].

\bibitem{GW}
P.H. Ginsparg and K.G. Wilson, \emph{A remnant of chiral symmetry on
  the lattice}, \prd{25}{1982}{2649}.

\bibitem{Has}
P.~Hasenfratz, \emph{Prospects for perfect actions},
\npps{63}{1998}{53} [\heplat{9709110}].

\bibitem{Luescher_abelian}
M.~L\"uscher, \emph{Topology and the axial anomaly in abelian lattice
  gauge theories}, \npb{538}{1999}{515} [\heplat{9808021}];
\emph{Abelian chiral gauge theories on the lattice with exact gauge
  invariance}, \npb{549}{1999}{295} [\heplat{9811032}];
\emph{Lattice regularization of chiral gauge theories to all orders of
  perturbation theory}, \jhep{06}{2000}{028} [\heplat{0006014}].

\bibitem{Golterman}
M.~Golterman, \emph{Lattice chiral gauge theories},
\npps{94}{2001}{189} [\heplat{0011027}].

\bibitem{parityanomaly}
A.N. Redlich, \emph{Gauge noninvariance and parity violation of
three-dimensional fermions}, \prl{52}{1984}{18};
\emph{Parity violation and gauge noninvariance of the effective gauge
  field action in three-dimensions}, \prd{29}{1984}{2366};\\
A.J. Niemi and G.W. Semenoff, \emph{Axial anomaly induced fermion
  fractionization and effective gauge theory actions in odd
  dimensional space-times}, \prl{51}{1983}{2077}.

\bibitem{ML}
M.~L\"uscher, \emph{Exact chiral symmetry on the lattice and the
  Ginsparg-Wilson relation}, \plb{428}{1998}{342} [\heplat{9802011}].

\bibitem{NPRL}
H.~Neuberger, \emph{A practical implementation of the overlap-Dirac
  operator}, \prl{81}{1998}{4060} [\heplat{9806025}].

\bibitem{Real}
H.~Suzuki, \emph{Real representation in chiral gauge theories on the
  lattice}, \jhep{10}{2000}{039} [\heplat{0009036}].

\bibitem{overlapDirac}
H.~Neuberger, \emph{Exactly massless quarks on the lattice},
\plb{417}{1998}{141} [\heplat{9707022}];
\emph{More about exactly massless quarks on the lattice},
\plb{427}{1998}{353} [\heplat{9801031}].

\bibitem{BS}
R.C. Brower and B.~Svetitsky, \emph{Hamiltonian domain wall fermions
  at strong coupling}, \prd{61}{2000}{114511} [\heplat{9912019}].

\bibitem{China}
W.~Bietenholz, \emph{Approximate Ginsparg-Wilson fermions for QCD}, in
 \emph{Non-perturbative methods and lattice QCD}, X.-Q. Luo and
 E.B. Gregory eds., World Scietific, Singapore 2001, p.~3
 [\heplat{0007017}].

\bibitem{equivalence}
R.~Narayanan, \emph{Ginsparg-Wilson relation and the overlap formula},
\prd{58}{1998}{097501} [\heplat{9802018}].

\bibitem{IzubuchiNishimura}
T.~Izubuchi and J.~Nishimura, \emph{Translational anomaly in chiral
  gauge theories on a torus and the overlap formalism},
  \jhep{10}{1999}{002} [\heplat{9903008}].

\bibitem{NarayananNishimura}
R.~Narayanan and J.~Nishimura, \emph{Parity-invariant lattice
  regularization of a three-dimensional gauge-fermion system},
  \npb{508}{1997}{371} [\hepth{9703109}].

\bibitem{KikukawaNeuberger}
Y.~Kikukawa and H.~Neuberger, \emph{Overlap in odd dimensions},
\npb{513}{1998}{735} [\heplat{9707016}].

\bibitem{KikukawaYamada}
Y.~Kikukawa and A.~Yamada, \emph{A note on the exact lattice chiral
  symmetry in the overlap formalism}, \heplat{9810024}.

\bibitem{CosteLuescher}
A.~Coste and M.~L\"uscher, \emph{Parity anomaly and fermion boson
  transmutation in three-dimensional lattice QED},
  \npb{323}{1989}{631}.

\bibitem{So}
H.~So, \emph{Induced Chern-Simons class with lattice fermions},
\ptp{73}{1985}{528};
\emph{Induced topological invariants by lattice fermions in odd
dimensions}, \ptp{74}{1985}{585}.

\bibitem{EPJC}
W.~Bietenholz, \emph{Solutions of the Ginsparg-Wilson relation and
  improved domain wall fermions}, \epjc{6}{1999}{537}
  [\heplat{9803023}].

\bibitem{Hasenfratz:1998ri}
P.~Hasenfratz, V.~Laliena and F.~Niedermayer, \emph{The index theorem
  in QCD with a finite cut-off}, \plb{427}{1998}{125}
  [\heplat{9801021}];\\
P.~Hasenfratz, \emph{Lattice QCD without tuning, mixing and current
renormalization}, \npb{525}{1998}{401} [\heplat{9802007}].

\bibitem{WBUJW}
W.~Bietenholz and U.-J. Wiese, \emph{Perfect lattice actions for quarks
  and gluons}, \npb{464}{1996}{319} [\heplat{9510026}].

\bibitem{BDS}
F. Berruto, M.C. Diamantini and P. Sodano,
\emph{On pure lattice Chern-Simons gauge theories},
\plb{487}{2000}{366} [\hepth{0004203}].

\bibitem{AK}
T.~Aoyama and Y.~Kikukawa, \emph{A lattice implementation of the
  eta-invariant and effective action for chiral fermions on the
  lattice}, \heplat{9905003}.

\bibitem{Fujikawa}
K.~Fujikawa, \emph{Path integral measure for gauge invariant fermion
  theories}, \prl{42}{1979}{1195};
\emph{Path integral for gauge theories with fermions},
\prd{21}{1980}{2848}.

\bibitem{nonlocalparity}
M.L. Ciccolini, C.D. Fosco and F.A. Schaposnik, \emph{Non local parity
  transformations and anomalies}, \plb{492}{2000}{214}
  [\hepth{0006248}].

\bibitem{3dsusy}
N.~Maru and J.~Nishimura, \emph{Lattice formulation of supersymmetric
  Yang-Mills theories without fine-tuning}, \ijmpa{13}{1998}{2841}
  [\hepth{9705152}];\\
J.~Nishimura, \emph{Applications of the overlap formalism to super
  Yang-Mills theories}, \npps{63}{1998}{721} [\heplat{9709112}].

\bibitem{CPTanomaly}
F.R. Klinkhamer, \emph{A CPT anomaly}, \npb{578}{2000}{277}
[\hepth{9912169}].

\bibitem{KlinkNishi}
F.R. Klinkhamer and J.~Nishimura, \emph{CPT anomaly in two-dimensional
  chiral ${\rm U}(1)$ gauge theories}, \prd{63}{2001}{097701}
  [\hepth{0006154}].

\end{thebibliography}
\end{document}